 \documentclass[12pt]{article}
{\baselineskip 12pt}  
\begin{document}
\title{
\vspace*{-20pt} {\bf Studying the Bell--Steinberger relation}}
\author{K. Urbanowski\footnote{e--mail:
K.Urbanowski@if.uz.zgora.pl; K.Urbanowski@proton.if.uz.zgora.pl} \\
University of Zielona Gora, Institute of Physics, \\
ul. Podgorna 50, 65--246 Zielona Gora, Poland.} \maketitle
{\noindent}{\em PACS numbers:}  03.65.Ca., 11.30.Er.,
11.10.St., 13.20.Eb., 14.40.Aq. 

\begin{abstract}
The Bell--Steinberger relation is analyzed. The questionable
points of the standard derivation of this relation are discussed.
It is shown that the use of a more accurate approximation than the
one usually used in the derivation of this relation can lead to
corrections to the right hand side of the standard
Bell--Steinberger relation.
\end{abstract}

\section{Introduction}

The Bell--Steinberger (BS) unitary relation \cite{Bell,Horwitz} is
considered as very useful and effective tool in searching for
properties of the $K_{0}, {\overline{K}}_{0}$ subsystem
\cite{Barmin} --- \cite{Hayakawa}. Some tests of  the fundamental
CPT and T invariance are based on the BS relation \cite{Barmin},
\cite{dafne1,dafne2}, \cite{Apostolakis} --- \cite{Hayakawa}. The
BS relation holds in the approximate Lee--Oehme--Yang (LOY) theory
of time evolution in the neutral kaon subsystem \cite{Lee1} ---
\cite{LOY1}, which follows from the Weisskopf--Wigner (WW)
approximation \cite{ww}. Khalfin \cite{leonid1,leonid2} has shown
that the BS relation in its original form is not true in the exact
theory. Similar conclusion can be drawn from the result contained
in \cite{plb-2002}. This means that the interpretation of the
results of all the test in neutral kaon subsystem, which are based
on the BS relation, can not be considered as ultimate. The proper
interpretation of such tests is impossible without a detailed
investigation of the weak points of this relation.

The original (standard) form of the BS relation is following,
\begin{equation}
\Big[ \frac{{\gamma}_{s} + {\gamma}_{l}}{2} - i(m_{s} - m_{l})
\Big] \, <s|l> = \sum_{F} <F|T|s>^{\ast}<F|T|l>. \label{BS-1}
\end{equation}
The derivation of this relation in such form is possible if the
transition operator $T$ exists \cite{Bell,Horwitz}. It is assumed
there that the operator $T$ describes transitions from states
belonging to the subspace, say ${\cal H}_{||}$, of states of
neutral kaons into the subspace  of their decay products, ${\cal
H}_{\perp}$. Here $|l>, \, |s> \, \in \, {\cal H}_{||}$, $|F> \,
\in \, {\cal H}_{\perp}$ and $\{|F> \}$ forms a complete
orthonormal set in ${\cal H}_{\perp}$. The Hilbert space  $\cal H
= {\cal H}_{||} \bigoplus {\cal H}_{\perp}$ is the state space of
the total system under consideration. Vectors $|l>, |s>$ are the
normalized eigenvectors of the effective Hamiltonian, $H_{||}$,
for the neutral $K$--mesons complex, for the eigenvalues
${\mu}_{l(s)} = m_{l(s)} - \frac{i}{2}{\gamma}_{l(s)}$
respectively,
\begin{equation}
H_{||} |l(s)> = {\mu}_{l(s)} |l(s)>. \label{H|l>=l|l>}
\end{equation}
There is
\begin{equation}
|l> = N_{l}(p_{l} |{\bf 1}> - q_{l}|{\bf 2}>), \;\; |s> =
N_{s}(p_{s}|{\bf 1}> + q_{s}|{\bf 2}>). \; \label{|l>,|s>}
\end{equation}
Here $|{\bf 1}>$ stands for vectors of the   $|K_{0}>,  \;
|B_{0}>$, $|n>$ --- neutrons, etc., type and $|{\bf 2}>$ denotes
antiparticles  of the particle "1": $|{\overline{K}}_{0}>, \;
|{\overline{B}}_{0}> \, |\overline{n} >$, and so on, $<{\bf
j}|{\bf k}> = {\delta}_{jk}$, $j,k =1,2$.

The  linear operator $H_{||}$ is the $(2 \times 2)$ nonhermitian
matrix,
\begin{equation}
H_{\parallel} \equiv \left(
\begin{array}{cc}
h_{11} & h_{12} \\
h_{21} & h_{22}
\end{array} \right) =
M - \frac{i}{2} \, \Gamma, \label{H-eff}
\end{equation}
(where $M=M^{+}$ is the mass matrix and
\begin{equation}
\Gamma = {\Gamma}^{+}\equiv i(H_{||} - H_{||}^{+}), \label{Gamma}
\end{equation}
denotes the decay matrix), acting in ${\cal H}_{||}$. Operators
$M$ and $\Gamma$ are linear. From (\ref{Gamma}) and (\ref{H-eff})
one finds that
\begin{equation}
{\Gamma}_{jk} = i (h_{jk} - h_{kj}^{\ast}), \; \; \; (j,k = 1,2).
\label{Gamma-jk}
\end{equation}

\section{Derivation of the standard Bell--Steinberger relation}

In deriving the relation (\ref{BS-1}) one usually invokes the
probability conservation \cite{Bell}, \cite{Horwitz}. The
probability conservation means that for every vector $|\psi ;t>
\in {\cal H}$ solving the Scr\"{o}dinger equation
\begin{equation}
i \frac{\partial}{\partial t} |\psi ;t> = H |\psi ;t>, \label{Sch}
\end{equation}
one has
\begin{equation}
||\; |\psi ; t> \;||^{2} = 1, \label{||=1}
\end{equation}
for every $t$. In the case considered the condition (\ref{||=1})
can be rewritten as follows
\begin{equation}
||\; |\psi ;t>_{||}\;||^{2} \;+ \; ||\;|\psi ;t>_{\perp}\;||^{2}
\; = \; 1, \label{psi||+psi-perp=1}
\end{equation}
where
\begin{equation}
|\psi ;t>_{||} \stackrel{\rm def}{=} P|\psi ; t>  \in {\cal
H}_{||}, \;\; |\psi ;t>_{\perp} \stackrel{\rm def}{=} Q|\psi ; t>
\in {\cal H}_{\perp}, \label{psi||}
\end{equation}
and $P$ is the projection operator onto the subspace ${\cal
H}_{||}$:
\begin{equation}
P \equiv |{\bf 1}><{\bf 1}| + |{\bf 2}><{\bf 2}|, \label{P}
\end{equation}
$Q$ is the projection operator onto the subspace of decay products
${\cal H}_{\perp}$, $Q \equiv I - P$.

The initial condition for the Eq. (\ref{Sch}) for the problem
considered is,
\begin{equation}
|\psi > \stackrel{\rm def}{=} |\psi ; t = 0> \equiv P|\psi>
\stackrel{\rm def}{=} |\psi >_{||} \in {\cal H}_{||}. \label{init}
\end{equation}

From (\ref{psi||+psi-perp=1}) it follows that
\begin{equation}
- \, \frac{\partial}{\partial t}||\; |\psi ;t>_{||}\;||^{2}\; =\;
\frac{\partial}{\partial t}||\;|\psi ;t>_{\perp}\;||^{2},
\label{||=-perp}
\end{equation}
Using this relation one usually assumes that its right hand side
can be identified with the following expression
\begin{equation}
\frac{\partial}{\partial t}||\;|\psi ;t>_{\perp}\;||^{2}\;
\stackrel{(?)}{=} \;\sum_{F} (<F|T|\psi
;t>_{||})^{\ast}\,<F|T|\psi ;t>_{||}, \label{psi-perp=}
\end{equation}
and thus, without proving if (or when) the relation (\ref
{psi-perp=}) is true, one finds that
\begin{equation}
- \, \frac{\partial}{\partial t}||\; |\psi ;t>_{||}\;||^{2}\;
\equiv \; \sum_{F} (<F|T|\psi ;t>_{||})^{\ast}\,<F|T|\psi
;t>_{||}, \label{BS-?}
\end{equation}
which leads to original BS formula (\ref{BS-1}). Indeed, following
\cite{Bell} and inserting
\begin{equation}
|\psi ;t >_{||}\; =\; x\,e^{\textstyle - i t{\mu}_{l}} |l> \;+ \;
y\,e^{\textstyle -it{\mu}_{s}} |s>, \label{xy}
\end{equation}
into (\ref{BS-?}), (where $x,y$ are arbitrary time--independent
number coefficients), and differentiating with respect to $t$ the
left hand side of the relation (\ref{BS-?}) and then putting
$t=0$, one obtains all the relations derived in \cite{Bell}. Of
course such a derivation is possible if ${\mu}_{l}, {\mu}_{s}$ do
not depend on time $t$.

Note that such a method of derivation of the BS relation
(\ref{BS-1}) can not be considered as  rigorous and correct.
Namely, it requires the existence of the transition operator $T$
for the particles under study. Next, the  decay of the states
under investigation must be described by the exponential function
of time $t$ for all times $t$. Such a assumption is not consistent
with the fundamental properties of quantum evolution. From the
properties of solutions of the Schr\"{o}dinger equation  and from
the basic principles of quantum theory it follows that the decay
process can not be exponential for times $t \rightarrow 0$ and for
times $t\rightarrow \infty$ \cite{Khalfin}. What is more, in
\cite{ur98} it was shown that the CPT Theorem of axiomatic quantum
field theory is not valid in the system containing exponentially
decaying particles. This means simply that the BS relation
(\ref{BS-1}) derived in such a way can not be used for designing
CPT violation tests.

The last weak point of this derivation of the relation
(\ref{BS-1}) is the following inconsistency. Using the
Schr\"{o}dinger equation it is not difficult to verify that
\begin{equation}
\frac{\partial}{\partial t}||\; |\psi ;t>_{||}\;||^{2}
\begin{array}[t]{l} \vline \,
\\ \vline \, {\scriptstyle t = 0 } \end{array}
\; \equiv \;0, \label{t=0}
\end{equation}
for the arbitrary initial condition $|\psi >  \in {\cal H}$. Thus
the relation (\ref{BS-?}) takes the following form at $t=0$,
\begin{equation}
-\,\frac{\partial}{\partial t}||\; |\psi ;t>_{||}\;||^{2}
\begin{array}[t]{l} \vline \,
\\ \vline \, {\scriptstyle t = 0 } \end{array}
\; \equiv \;0 \;= \; \sum_{F} (<F|T|\psi>_{||})^{\ast}\,<F|T|\psi
>_{||} \, \neq 0.
\label{neq-0}
\end{equation}
So the left hand side of the relation (\ref{BS-1}) equals zero
whereas the right hand side of this relation is non zero at $t=0$.
The conclusion is that one should be very careful using the
original BS relation (\ref{BS-1}) as a tool for searching for
properties of the neutral kaon and similar complexes.

In the original form of the BS relation, \cite{Bell}
--- \cite{Hayakawa}, the vectors $|l>$ and $|s>$ are
understood as the eigenvectors for the LOY effective Hamiltonian,
$H_{LOY}$. That is $H_{||} \equiv H_{LOY}$ in such a case.

\section{Approximate effective Hamiltonians and the BS relation}

In many papers the observation was made that in order to obtain
the left hand side of the BS relation (\ref{BS-1}) one need not
use the method based on the relations (\ref{psi-perp=}) ---
(\ref{xy}). It appears that the equivalent relation can be derived
directly form the eigenvalue equation (\ref{H|l>=l|l>}), (see,
eg., \cite{Tsai1,Tsai2,Tsai3,Lee2}\,). Indeed directly from
(\ref{H|l>=l|l>}) one finds
\begin{equation}
\Big[ \frac{{\gamma}_{s} + {\gamma}_{l}}{2} - i(m_{s} - m_{l})
\Big] \, <s|l> = <s|\Gamma |l>, \label{BS-2}
\end{equation}
which within the LOY approximation is equivalent to (\ref{BS-1}).

This method of the derivation of the BS relation is free of the
above mentioned inconsistencies. It has an advantage over the
original one \cite{Bell,Horwitz} because one does not make use of
the transition matrix $T$. Simply, the assumption about the
existence of the $T$ operator is unnecessary in this case. It is a
very important property of this method because, in fact, the
correct definition of the scattering matrix, $S \equiv I + iT$,
and thus the $T$--matrix, makes use of asymptotic states. Such
states do not exist for unstable particles and $K_{0},
{\overline{K}}_{0}$ mesons are unstable. What is more within this
method one need not assume that the decay is exponential.

The accuracy of the relation (\ref{BS-2}) is determined by the
accuracy of the approximation leading to the $H_{||}$ used there.
If one inserts to the eigenvalue equation (\ref{H|l>=l|l>}) the
effective Hamiltonian $H_{||} \equiv H_{LOY}$ then one comes to
the picture equivalent to the original BS treatment of this
problem. On the other hand, if one uses the exact effective
Hamiltonian $H_{||}$ then the relation (\ref{BS-2}) will not
describe the approximate but it will describe the real properties
of the system under consideration. The use of a more accurate
approximation for $H_{||}$ than the LOY approximation in
(\ref{H|l>=l|l>}), and thus in (\ref{BS-2}),  will lead to a
description of the system considered, which can be sensitive to
possible effects unreachable by means of  the LOY method.

The LOY effective Hamiltonian, $H_{LOY}$, can be expressed in a
compact form as follows \cite{improved}
\begin{equation}
H_{LOY} = m_{0}P - \Sigma (m_{0}) = M^{LOY} - \frac{i}{2}
{\Gamma}^{LOY}, \label{H-LOY}
\end{equation}
where
\begin{eqnarray}
\Sigma ( \epsilon ) & = & PHQ \frac{1}{QHQ - \epsilon - i 0} QHP
\nonumber \\&=& {\Sigma}^{R}(\epsilon) + i {\Sigma}^{I}(\epsilon),
\label{Sigma}
\end{eqnarray}
and ${\Sigma}^{R}(\epsilon = {\epsilon}^{\ast}) =
{\Sigma}^{R}(\epsilon = {\epsilon}^{\ast})^{+}$,
${\Sigma}^{I}(\epsilon = {\epsilon}^{\ast})$
$={\Sigma}^{I}(\epsilon = {\epsilon}^{\ast})^{+}$. The operator
${\Sigma}^{I}(\epsilon)$ we are especially interested in has the
following form
\begin{equation}
{\Sigma}^{I}(\epsilon) \equiv \pi \, PHQ \, \delta (QHQ - \epsilon
) \, QHP. \label{Sigma-I}
\end{equation}

Within the LOY approach vectors $|{\bf 1}>$, $|{\bf 2}>$ are
normalized eigenstates of the free Hamiltonian, $H^{(0)}$, (here
$H \equiv H^{(0)} + H^{I}$  is the total Hamiltonian of the system
considered), for 2-fold degenerate eigenvalue $m_{0}$:
\begin{equation}
H^{(0)} |{\bf j} > = m_{0} |{\bf j }>, \; \;  (j = 1,2).
\label{H-0-j}
\end{equation}
$H^{I}$ denotes the interaction which is responsible for the decay
process.

From (\ref{H-LOY}) one finds that
\begin{equation}
{\Gamma}^{LOY} = 2 {\Sigma}^{I}(m_{0}), \label{Gamma-LOY}
\end{equation}
which means that
\begin{eqnarray}
{\Gamma}^{LOY}_{jk} &=& 2 \pi <{\bf j}|HQ \delta (QHQ - m_{0})
QH|{\bf k}>,   \label{Gamma-LOY-jk} \\
& \equiv & \pi \sum_{F} \delta (E_{F} - m_{0})<{\bf
j}|PH|F><F|HP|{\bf k}>,    \label{Gamma-LOY-jk-F}
\end{eqnarray}
where $(j,k = 1,2)$,  $H|F> = E_{F}|F>$ and $ \sum_{F} |F><F|
\equiv Q$.  So within the LOY approximation using
(\ref{Gamma-LOY-jk-F}) one finds that in the CPT invariant system,
the right hand side of the relation (\ref{BS-2}) takes the
following form
\begin{eqnarray}
<s|{\Gamma}^{LOY} |l> &=& 2 <s| {\Sigma}^{I}(m_{0}) |l>
\label{B-G-LOY-s-l} \\
&\equiv & 2 \pi \sum_{F} \delta (E_{F} - m_{0})
<F|HP|s>^{\ast}\;<F|HP|l>. \label{B-G-LOY-sl-F}
\end{eqnarray}
This is the standard picture which one meets in the literature.
Note that this expression coincides with the right hand side of
(\ref{BS-1}).

Now, if one uses the exact $H_{||}$ instead of the approximate
$H_{LOY}$  in the relation (\ref{BS-2}) then one can expect that
the right hand side of the BS relation (\ref{BS-2}) will differ
from (\ref{B-G-LOY-sl-F}). The exact effective Hamiltonian
$H_{||}$ is time dependent \cite{Horwitz,plb-2002}, \cite{ur98}
--- \cite{ijmpa95}, $H_{||} = H_{||}(t)$, in the nontrivial case,
and can be expressed as follows \cite{plb-2002}, \cite{ur98} ---
\cite{ijmpa95}
\begin{equation}
H_{||} = PHP + V_{||}(t), \label{H-||-(t)}
\end{equation}
where the nonhernmitian operator $V_{||}(t)$ has the following
property
\begin{equation}
V_{||}(t=0) \equiv 0. \ \label{V-t=0}
\end{equation}
(The "nontrivial case" is understood here as the $[P,H] \neq 0$
case). In the nontrivial case the property that the effective
Hamiltonian depends on time, $H_{||}= H_{||}(t)$, has the
following consequence: In CPT invariant but CP noninvariant system
the diagonal matrix elements $h_{jj}$, $(j=1,2)$, can not be equal
for $t>0$ \cite{plb-2002} and all coefficients $p_{s}, p_{l},
q_{s}, \ldots$ appearing in the formulae (\ref{|l>,|s>}) are
different and time dependent \cite{leonid1,leonid2,ijmpa93}. The
same is true about the eigenvalues ${\mu}_{l},{\mu}_{s}$.

From (\ref{H-||-(t)}) it follows that the matrix $\Gamma$ can be
expressed as follows
\begin{equation}
\Gamma \equiv \Gamma (t) = i \Big(V_{||}(t) - (V_{||}(t))^{+}
\Big). \label{Gamma(t)}
\end{equation}
So, the relation (\ref{V-t=0}) means that in the exact case
$\Gamma (t=0) = 0$. From this property and from properties of the
eigenvectors $|l> = |l^{t}>$ and $|s> = |s^{t}>$
\cite{ijmpa93,ijmpa95} one concludes that at the initial instant
$t = 0$ the BS relation (\ref{BS-2}) becomes trivial: 0 = 0. It
contrasts with the relation (\ref{neq-0}) and it is consistent
with the basic assumptions of quantum theory. This is simplest
general conclusion which can be obtained for the exact case. For
the considered models of interactions leading to the decay of
$K_{0}, {\overline{K}}_{0}$ mesons it is practically impossible to
calculate the exact effective Hamiltonian $H_{||}$. Nevertheless,
one can study the BS relation using the more accurate approximate
effective Hamiltonians $H_{||}$ than $H_{LOY}$ and thus one can
look for possible deviations from the standard (i.e., LOY)
picture. An example of the more accurate $H_{||}$ that $H_{LOY}$
is given in \cite{ur98,ijmpa93,improved,Pi00,Pi03}.

The approximate formulae for $H_{\parallel}(t)$ have been  derived
there using  the  Krolikowski--Rzewuski (KR) equation   for the
projection of a state vector \cite{KR}, which results from the
Schr\"{o}dinger equation (\ref{Sch}) for  the  total system under
consideration, and, in the  case  of initial conditions of the
type (\ref{init}), takes the following form
\begin{equation}
( i \frac{\partial}{ {\partial} t} - PHP )
U_{\parallel}(t)|\psi>_{||}
 =  - i \int_{0}^{\infty} K(t - \tau ) U_{\parallel}
( \tau )|\psi>_{||} d \tau,   \label{KR1}
\end{equation}
where $ U_{\parallel} (0)  =  P$, and $U_{||}(t)$ is the evolution
operator for the subspace ${\cal H}_{||}$,
\begin{equation}
K(t)  =  {\mit \Theta} (t) PHQ \exp (-itQHQ)QHP, \label{K}
\end{equation}
and ${\mit \Theta} (t)  =  { \{ } 1 \;{\rm for} \; t \geq 0, \; \;
0 \; {\rm for} \; t < 0 { \} }$.

The integro--differential equation (\ref{KR1}) is equivalent to
the following differential one (see \cite{bull} --- \cite{KR})
\begin{equation}
( i \frac{\partial}{ {\partial} t} - H_{||}(t) )
U_{\parallel}(t)|\psi>_{||} = 0, \label{KR2}
\end{equation}
where the effective Hamiltonian  $H_{||}(t)$ has the form
(\ref{H-||-(t)}). Taking into account (\ref{KR1}), (\ref{KR2}) and
(\ref{H-||-(t)}) one finds from (\ref{KR1}))
\begin{equation}
V_{\parallel} (t) U_{\parallel} (t) = - i \int_{0}^{\infty} K(t -
\tau ) U_{\parallel} ( \tau ) d \tau. \label{V||=def}
\end{equation}
This relation can be used to obtain the approximate formula for
$V_{||}(t)$. From (\ref{V||=def}) one fins to the lowest
nontrivial order \cite{ijmpa93,pra}
\begin{equation}
V_{\parallel}(t) \cong V_{\parallel}^{(1)} (t) \stackrel{\rm
def}{=} -i \int_{0}^{\infty} K(t - \tau ) \exp {[} i ( t - \tau )
PHP {]} d \tau . \label{V||=approx}
\end{equation}
The use of $P$ defined by the relation (\ref{P}) leads to
\cite{improved,Pi00}
\begin{eqnarray}
V_{||}(t) & = & - \frac{1}{2} \Xi (H_{0} + \kappa;t)\, \Big[ \Big(
1 - \frac{H_{0}}{\kappa} \Big)P + \frac{1}{\kappa} PHP\Big]
\nonumber \\
&& - \frac{1}{2} \Xi (H_{0} - \kappa;t)\, \Big[ \Big( 1 +
\frac{H_{0}}{\kappa} \Big)P - \frac{1}{\kappa} PHP\Big],
\label{V||(t)}
\end{eqnarray}
where
\begin{equation}
H_{0} = \frac{1}{2} (H_{11} + H_{22}), \;\;\;\; \kappa =
\sqrt{|H_{12}|^{2} + \frac{1}{4}(H_{11} - H_{22})^{2}},
\label{kappa}
\end{equation}
and
\begin{equation}
H_{jk} = <{\bf j}|H|{\bf k}>, \; \;\;\;\; (j,k = 1,2),
\label{H-jk}
\end{equation}
\begin{equation}
\Xi (x;t) \stackrel{\rm def}{=} PHQ \frac{e^{\textstyle{- it (QHQ
- x)}} - 1}{QHQ - x} QHP, \label{Xi}
\end{equation}
Note that $V_{\parallel}(t) \cong V_{\parallel}^{(1)} (t) = 0$ at
$t=0$, which agrees with the general property of the exact
$H_{||}(t)$ and $V_{||}(t)$ (see (\ref{V-t=0})). The expression
(\ref{V||(t)}) leads by (\ref{Gamma(t)})  to a very complicated
form of $\Gamma (t)$. Such a form of ${\Gamma}$ is very hard  to
compare it with ${\Gamma}^{LOY}$ and thus to relate it to the
right hand side of the original BS relation (\ref{BS-1}). The form
(\ref{V||(t)}) of $V_{||}(t)$ becomes simpler when
\begin{equation}
\kappa \ll H_{0}, \label{kappa<}
\end{equation}
because then
\begin{equation}
\Xi ( H_{0} \pm \kappa ; t)\, \simeq \,\Xi (H_{0} ;t)\, \pm \,
\kappa \, \frac{\partial \Xi (x;t)}{\partial x}\begin{array}[t]{l}
\vline \,
\\ \vline \, {\scriptstyle x = H_{0} } \end{array} +
\ldots \; . \label{Xi-kappa<}
\end{equation}
So, if the condition (\ref{kappa<}) holds then
\begin{equation}
V_{||}(t) \,\simeq \, - \,\Xi (H_{0};t) \, - \, \frac{\partial \Xi
(x;t)}{\partial x}\begin{array}[t]{l} \vline \,
\\ \vline \, {\scriptstyle x = H_{0} } \end{array}
(PHP \, - \, H_{0}\, P)\;
 + \ldots \; . \label{V(t)-kappa<}
\end{equation}
(Note, that due to the presence of resonance terms the second term
on the right hand side of the above expression, that is
$\frac{\partial \Xi (x;t)}{\partial x}$, need not be small).  This
last expression is simpler than (\ref{V||(t)}) but it also leads
to time dependent $\Gamma$ and thus $<s|\Gamma (t) |l>$. Such
$<s|\Gamma (t) |l>$ can not be compared with the BS relation
(\ref{BS-1}), which is applied for asymptotic times $t \rightarrow
\infty$. One needs $V_{||}(t)$ for times $t$ which are at least of
order of the lifetimes, ${\tau}_{l}, {\tau}_{s}$ for states $|l>,
|s>$, that is for $t \sim {\tau}_{l}$. It can be achieved using
$V_{||} \stackrel{\rm def}{=} \lim_{t \rightarrow \infty}
V_{||}^{(1)}(t)$ instead of (\ref{V||(t)}). There is
\begin{equation}
\lim_{t \rightarrow \infty} \, \Xi (x;t) = \Sigma (x).
\label{Xi-lim}
\end{equation}
Thus
\begin{eqnarray}
V_{||} &\stackrel{\rm def}{=}& \lim_{t \rightarrow \infty}
V_{||}^{(1)}(t) \nonumber \\
& = & - \frac{1}{2} \Sigma (H_{0} + \kappa )\, \Big[ \Big( 1 -
\frac{H_{0}}{\kappa} \Big)P + \frac{1}{\kappa} PHP\Big]
\nonumber \\
&& - \frac{1}{2} \Sigma (H_{0} - \kappa )\, \Big[ \Big( 1 +
\frac{H_{0}}{\kappa} \Big)P - \frac{1}{\kappa} PHP\Big],
\label{V||}
\end{eqnarray}

To realize the purpose of this paper it is sufficient to consider
the case (\ref{kappa<}). So, if condition (\ref{kappa<}) holds
then
\begin{equation}
\Sigma ( H_{0} \pm \kappa )\, \simeq \,\Sigma (H_{0})\, \pm \,
\kappa \, \frac{\partial \Sigma(x)}{\partial x}\begin{array}[t]{l}
\vline \,
\\ \vline \, {\scriptstyle x = H_{0} } \end{array} +
\ldots \; . \label{Sigma-kappa<}
\end{equation}
which, by (\ref{V||}), yields
\begin{equation}
V_{||} \,\simeq \, - \,\Sigma (H_{0}) \, - \, \frac{\partial
\Sigma (x)}{\partial x}\begin{array}[t]{l} \vline \,
\\ \vline \, {\scriptstyle x = H_{0} } \end{array}
(PHP \, - \, H_{0}\, P)\;
 + \ldots \; . \label{V-kappa<}
\end{equation}
Thus, taking into account (\ref{Gamma-LOY}) and (\ref{H-||-(t)}),
(\ref{Gamma(t)}) one finds that if condition (\ref{kappa<}) takes
place then
\begin{equation}
\Gamma \, \equiv \, {\Gamma}^{(0)}\,+ \, \Delta \Gamma ,
\label{G+dG}
\end{equation}
where
\begin{equation}
{\Gamma}^{(0)}\, = \, 2 \, {\Sigma}^{I}(H_{0}), \label{G-0}
\end{equation}
and
\begin{eqnarray}
\Delta \Gamma & = & - \,i \Big[ \frac{\partial \Sigma(x)}{\partial
x}\begin{array}[t]{l} \vline \,
\\ \vline \, {\scriptstyle x = H_{0} } \end{array} (PHP \, - \,
H_{0} \, P) \nonumber \\
&& \;\;\;\; - \, (PHP \, - H_{0}\,P) \Big( \frac{\partial \Sigma
(x)}{\partial x} {\Big)}^{+} \begin{array}[t]{l} \vline \,
\\ \vline \, {\scriptstyle x = H_{0} } \end{array}
\Big]. \label{dG}
\end{eqnarray}
Thus in this case
\begin{equation}
<s| \Gamma |l> \, \simeq \, <s|{\Gamma}^{(0)}|l> \, + \, <s|
\Delta \Gamma |l>, \label{G-sl-dG}
\end{equation}
which evidently differs from (\ref{B-G-LOY-s-l}).

Note, that all the above discussed expressions (\ref{V||(t)}) ---
(\ref{dG}) have been derived without assuming any symmetries of
the type  CP--,  T--, or CPT--symmetry  for  the  total
Hamiltonian  H   of   the system considered. Now let us assume
that the CPT symmetry is conserved, that is
\begin{equation}
[ \Theta , H ] = 0, \label{[CPT,H]=0}
\end{equation}
where $\Theta$ is the antiunitary operator: $\Theta \stackrel{\rm
def}{=} {\cal C}{\cal P}{\cal T}$ and  $\cal C$ denotes the charge
conjugation, $\cal P$ --- space inversion (parity) and $\cal T$
--- time reversal transformations. Let us assume also that
the subspace of neutral kaons ${\cal H}_{\parallel}$ is  invariant
under $\Theta$:
\begin{equation}
[{\Theta}, P] = 0. \label{[CPT,P]=0}
\end{equation}
When these two last assumptions hold then $H_{11} = H_{22}, \;
\kappa \equiv |H_{12} |$ and $H_{0} \equiv H_{11} \equiv H_{22}$
and also ${\Sigma}_{11} ( \varepsilon = {\varepsilon}^{\ast} )
\equiv {\Sigma}_{22} ( \varepsilon = {\varepsilon}^{\ast} )
\stackrel{\rm def}{=} {\Sigma}_{0} ( \varepsilon =
{\varepsilon}^{\ast} )$. So, when the total system is CPT
invariant all the expressions for the approximate $V_{||}(t)$,
(\ref{V||(t)}), (\ref{V(t)-kappa<}), $V_{||}$, (\ref{V||}),
(\ref{V-kappa<}), and $\Gamma, \; {\Gamma}^{(0)}$, (\ref{G-0}),
(\ref{dG}), become simpler. It is very important that this
approximate $V_{||}$ leads to effective Hamiltonian $H_{||}$
possessing properties consistent with the properties of the exact
effective Hamiltonian: Analogously to the properties of the exact
effective Hamiltonian \cite{plb-2002}  its diagonal matrix
elements are not equal if the total system under considerations is
CPT invariant but CP nonivariant \cite{ur98,ijmpa93,ijmpa95}. This
property is absent in the LOY approximation and therefore the
approach basing on the LOY effective Hamiltonian is unable to
reflect correctly all the properties of the real system. So the
description of the properties of the $K_{0}, {\overline{K}}_{0}$
complex within the use of the above described approximation based
on the KR equation should lead to a more realistic picture of the
behavior of this complex than that given by the LOY and related
approaches.

In the case of preserved CPT symmetry, i.e., when conditions
(\ref{[CPT,H]=0}), (\ref{[CPT,P]=0}) hold, one can identify
$H_{0}$ appearing in (\ref{G-0}) with $m_{0}$ from the formula
(\ref{Gamma-LOY}): $H_{0} \equiv m_{0}$. This means that when the
total system preserves CPT symmetry and the condition
(\ref{kappa<}) holds then
\begin{equation}
{\Gamma}^{(0)} \, \equiv \, {\Gamma}^{LOY}, \label{G-0=G-LOY}
\end{equation}
Note that in this case still $\Delta \Gamma \neq 0$ (see
(\ref{dG})). Therefore one again finds that
\begin{eqnarray}
<s| \Gamma |l> \, &\simeq &\, <s|{\Gamma}^{(0)}|l> \, + \, <s|
\Delta \Gamma |l> \nonumber \\ &\equiv &\,
 <s|{\Gamma}^{LOY}|l> \, + \, <s|
\Delta \Gamma |l> \nonumber \\ &\neq &\,  <s| {\Gamma}^{LOY} |l>,
\label{G-LOY-sl-dG}
\end{eqnarray}

One observes from (\ref{V(t)-kappa<}), (\ref{V-kappa<}) (or
(\ref{dG})) that if the total Hamiltonian $H$ has the following
property
\begin{equation}
PHP \equiv H_{0}\, P, \label{P-H12=0}
\end{equation}
then simply
\begin{equation}
V_{||} = - \Sigma (H_{0}), \label{V-H-0}
\end{equation}
and $\Delta \Gamma \, \equiv \, 0$. So, in such a case
\begin{equation}
\Gamma \, \equiv \, {\Gamma}^{LOY}, \label{G=G-LOY}
\end{equation}
and $H_{||} \equiv H_{LOY}$.

The solution of the condition (\ref{P-H12=0}) is simple
\begin{equation}
H_{12} = H_{21} = 0.  \label{H12=0}
\end{equation}
This means (by (\ref{H-jk})) that if the first order $|\Delta S| =
2$ transitions do not occur in the $K_{0}, {\overline{K}}_{0}$
complex than the BS relation (\ref{BS-2}) with $\Gamma \equiv
{\Gamma}^{LOY}$ and with $\Gamma$ given by (\ref{Gamma(t)}),
(\ref{G+dG}) coincide. This also means the the original BS
relation should not be used for designing, eg., tests verifying
the existence of such interactions.

\section{Final remarks}

Note, that from the relation (\ref{B-G-LOY-sl-F}), using the
Schwartz inequality, the  conclusion of the following type is
drawn in the literature (see  \cite{Bell}, \cite{Lee2} and also
see (6.85) in \cite{Bigi}, etc.)
\begin{eqnarray}
|<s|\Gamma |l>|^{2} &\equiv & | \pi \sum_{F} \delta (E_{F} -
m_{0}) <F|HP|s>^{\ast}\;<F|HP|l>|^{2} \nonumber \\
&\leq &\pi\sum_{F} \delta (E_{F} - m_{0})
<F|HP|s>^{\ast}\;<F|HP|s> \times \nonumber \\
& \times &  \pi \sum_{F} \delta (E_{F} - m_{0})
<F|HP|l>^{\ast}\;<F|HP|l>. \label{Gs-Gl}
\end{eqnarray}
This inequality is interpreted as follows
\begin{equation}
|<s|\Gamma |l>|^{2} \, \leq \, {\Gamma}_{s} \, {\Gamma}_{l},
\label{Gs-Gl>}
\end{equation}
where the decay widths ${\Gamma}_{s}, {\Gamma}_{l}$ are identified
with
\begin{equation}
{\Gamma}_{l(s)} \, = \, \pi \sum_{F} \delta (E_{F} - m_{0})
<F|HP|l(s)>^{\ast}\;<F|HP|l(s)>. \label{Gs(l)}
\end{equation}
The inequality (\ref{Gs-Gl>}) together with the BS relation
(\ref{BS-2}) is used, eg., to estimate the product $|<s|l>|$ (see
e.g., \cite{Bell}, \cite{Bigi}, \cite{Lee2}). In the light of the
relations (\ref{G+dG}), (\ref{G-LOY-sl-dG}) such estimations and
similar conclusions can be considered as consistent with the real
properties of the system under investigation only if it were
$\Gamma \equiv {\Gamma}^{LOY}$.

Note that the inequality (\ref{Gs-Gl}) need not be true in the
case, when the relation (\ref{G+dG}) (or when the earlier
expressions for $V_{||}$)  take place. Then simply $\Gamma \neq
{\Gamma}^{LOY}$. This means that the estimations of type
(\ref{Gs-Gl>}) need not be true in such a case. So, keeping in
mind relations (\ref{G+dG}), (\ref{G-LOY-sl-dG}) one should be
very careful while considering the tests performed in $K_{0},
{\overline{K}}_{0}$ complex within the use of the original BS
relations (\ref{BS-1}) (or (\ref{BS-2}), where $\Gamma =
{\Gamma}^{LOY}$) as the crucial one.

On the other hand, due to the properties of the matrix $\Gamma$
($\Gamma$ is the linear and hermitian matrix) the expression
expression $<s|\Gamma |l>$ defines the hermitian form in the
subspace ${\cal H}_{||}$. Indeed, for every $|\phi >, |\psi > \,
\in \, {\cal H}_{||}$ one can define
\begin{equation}
(\phi ,\psi) \stackrel{\rm def}{=} <\psi |\Gamma |\phi >.
\label{(f|p)}
\end{equation}
Now, if the matrix $\Gamma$ is positive definite then the form
(\ref{(f|p)}) is a positive definite hermitian form. It is not
difficult to verify that the  form $(\phi ,\psi)$ must then
fulfill all the requirements of the scalar product. Therefore in
this case for the product $(\phi ,\psi )$ the Schwartz inequality
holds, which reads
\begin{equation}
|(\phi ,\psi )|^{2} \, \leq \, (\phi ,\phi )\,(\psi ,\psi ),
\label{Schwartz-1}
\end{equation}
Within the use of the definition (\ref{(f|p)}) this inequality can
be rewritten as follows
\begin{equation}
|<\psi |\Gamma |\phi >|^{2} \;\; \leq \;\;<\psi |\Gamma |\psi >
\,<\phi |\Gamma |\phi >. \label{Schwartz-2}
\end{equation}
This inequality is true for every $|\psi
>, |\phi> \in {\cal H}_{||}$ only if $\Gamma$ is positive definite.

Now if the eigenvectors $|s>, |l>$ of $H_{||}$ are inserted into
(\ref{Schwartz-2}) then one can find that
\begin{equation}
|<s|\Gamma |l >|^{2} \, \leq \,<s |\Gamma |s > \,<l |\Gamma |l>.
\label{SCHWARTZ}
\end{equation}
Thus, using the eigenvalue equation (\ref{H|l>=l|l>}) for $H_{||}$
and the relation (\ref{Gamma}) one can conclude that
\begin{equation}
|<s|\Gamma |l >|^{2} \, \leq \, {\gamma}_{s} \, {\gamma}_{l}.
\label{gs-gl>}
\end{equation}
One should stress on that ${\gamma}_{s}, {\gamma}_{l}$ appearing
in this inequality are determined by the solutions of the
eigenvalue problem for $H_{||}$, but not by the relations
(\ref{Gamma-LOY-jk}), (\ref{Gamma-LOY-jk-F}) or (\ref{Gs(l)}).

Unfortunately there has not been published any rigorous proof that
the exact $\Gamma$ should be positive definite. So the inequality
(\ref{Schwartz-1}) (and therefore the inequalities
(\ref{Schwartz-2}) --- (\ref{gs-gl>})) can not be considered as
definitely valid. What is more there exists reasons leading to the
conclusion that the matrix $\Delta \Gamma$ defined by the formula
(\ref{dG}) is not positive and thus the matrix $\Gamma$ connected
with $\Delta \Gamma$ by the relation (\ref{G+dG}) need not be
positive definite. Such a conclusion follows from the generalized
Fridrichs--Lee model (see \cite{chiu}) calculations performed in
\cite{ijmpa93} and \cite{ur98}. Results obtained there lead  to
the following form of $\Gamma = {\Gamma}^{FL}$ in the CPT
invariant case,
\begin{equation}
{\Gamma}^{FL} \equiv \left(
\begin{array}{cc}
{\gamma}_{11} & {\gamma}_{12} \\
{\gamma}_{21} & {\gamma}_{22}
\end{array} \right) = {\Gamma}^{(0)} + \Delta \gamma,
\label{G^FL}
\end{equation}
where ${\Gamma}^{(0)} = {\Gamma}^{LOY}$, ${\gamma}_{jk} =
{\Gamma}^{LOY}_{jk} + \Delta {\gamma}_{jk}, \;(j,k=1,2) $ and
\begin{equation}
\Delta {\gamma}_{11} = \Delta {\gamma}_{22} = - \frac{1}{2} \,
\frac{\Re \, (m_{21} {\Gamma}_{12}^{LOY})}{m_{0} - \mu},
\label{dg-00}
\end{equation}
\begin{equation}
\Delta {\gamma}_{12} = (\Delta {\gamma}_{21})^{\ast} = -
\frac{1}{4} \frac{m_{12}}{m_{0} - \mu} ({\Gamma}_{11}^{LOY} +
{\Gamma}_{22}^{LOY} ), \label{dg-12}
\end{equation}
$\Re \, (z)$ denotes the real part of $z$, $m_{jk} \equiv H_{jk}$,
($j,k =1,2$), $m_{0} \equiv H_{11} = H_{22}$ and  $\mu$ denotes
the mass of the decay products. The last formula  was obtained
assuming that $|m_{12}|\equiv |H_{12}| \ll (m_{0}- \mu )$.

From the Sylvester Theorem it follows that matrix $\Delta \gamma$
can be positive definite only and only if there are: ${\Delta
\gamma}_{11}> 0$ and $\det \Delta \gamma >0$. There is,
\begin{equation}
\det \Delta \gamma = \frac{1}{16} \frac{1}{(m_{0} - \mu)^{2}}
\Big[ 4\, \Big(\Re \, (m_{21}\,{\Gamma}^{LOY}_{12}){\Big)}^{2} -
|m_{12}|^{2} ({\Gamma}_{11}^{LOY} + {\Gamma}_{22}^{LOY} )^{2}
\Big], \label{det-dg}
\end{equation}
in the case considered. Now if we assume that ${\Gamma}^{LOY}$ is
positive definite, which is is equivalent to the assumptions that
$|{\Gamma}_{12}^{LOY}|^{2} \,\leq
\,{\Gamma}_{11}^{LOY}\,{\Gamma}_{22}^{LOY}$ and
${\Gamma}_{11}^{LOY} > 0$, then $\det \, \Delta \, \gamma \, \leq
\, 0$. So, if in  this case  ${\Gamma}^{(0)} = {\Gamma}^{LOY}$ is
positive definite then the matrix $\Delta \gamma$ can not be
positive definite. Therefore the matrix ${\Gamma}^{FL} =
{\Gamma}^{(0)} + \Delta \gamma$ need not be positive definite.
This means that in such a model the hermitian form (\ref{(f|p)})
need not fulfill the requirements of the scalar product and thus
inequalities of type (\ref{Schwartz-1})
--- (\ref{gs-gl>}) can not be considered as definitely valid.

Similar considerations lead to the conclusion that in the general
case (\ref{dG}),  the matrix $\Delta \Gamma$ need not be positive.
Thus inequalities of type (\ref{Schwartz-1}) --- (\ref{gs-gl>})
may not be valid in the case of the relation (\ref{G+dG}).

From the above considerations the following conclusions follow: If
searching for the properties of neutral kaon and similar complexes
one is going to use the estimations of type (\ref{gs-gl>}), one
always should verify whether the matrix $\Gamma$ is positive
definite or not. The inequality (\ref{gs-gl>}) is true only for
positive definite $\Gamma$. Of courses, one always expects and
assumes that $\Gamma$ should be positive defined. Nevertheless we
should remember that our assumptions or expectations can never
replace an inspection or a rigorous proof.

From the relations (\ref{G^FL}) --- (\ref{det-dg}) it follows that
if there exist interactions in the system considered leading to
the matrix elements $<{\bf 1}|H|{\bf 2}> \neq 0$, that is, if
there exist interactions allowing the first order $|\Delta S| =2$
transitions, then the matrix $\Gamma$ calculated within the more
accurate approximation than $H_{LOY}$ need not be positive
definite. This means that in such a case conclusions following
from the inequality of type (\ref{gs-gl>}) can not be considered
as ultimate. The same concerns tests for the existence of such
interactions based on this inequality.

The standard derivation of the BS relation makes essential use of
the assumption about the exponential form of the decay law of the
neutral kaons. As it was mentioned earlier such an assumption (see
\cite{ur98}) and other inconsistencies of this derivation cause
that all test CPT invariance basing on the BS relation
(\ref{BS-1}) can not be considered as crucial.

Note that the BS relation (\ref{BS-2}) and the inequality
(\ref{gs-gl>}) only contains quantities appearing in the
eigenvalue equations (\ref{H|l>=l|l>}) for the effective
Hamiltonian $H_{||}$. Therefore one can assume that real
properties of subsystems considered (like the neutral kaon
complex, etc.), will be described by solutions of the eigenvalue
problem for the exact $H_{||}$. In other words, the result
(\ref{gs-gl>}) means that one can expect what follows: Estimations
of parameters describing the neutral kaon complex performed within
the use of the BS relation (\ref{BS-2}) and the inequality
(\ref{gs-gl>}) describe real properties of this complex only if
the quantities, which one inserts there, are extracted directly
from the experiments and the positivity of $\Gamma$ is rigorously
proved. Of course these experiments must be designed in such a
manner that the interpretation of  the results of these tests is
independent of the approximation used to describe the system under
investigations. This means that, eg., the parameters of the type
$\gamma_{s}, \gamma_{l}$ or ${\Gamma}_{l(s)}$, $\Gamma_{jk}$, can
not be extracted using the relations of the type (\ref{Gs(l)}). If
one is unable to realize the test in such a manner then the
interpretation of its results basing on the BS relation
(\ref{BS-2}) need not reflect the real properties of the system
under investigation. There is the following reason for such a
conclusion. Simply, comparing the form of the formulae
(\ref{Gamma-LOY-jk}), (\ref{Gamma-LOY-jk-F}), (\ref{Gs(l)}) with
expressions (\ref{V||}), (\ref{G+dG}) and (\ref{dG}) obtained
within the more accurate approximation one finds that the real
structure of the processes and interactions in the subsystem under
investigations can be more complicated than it follows from the
standard formulae (\ref{Gamma-LOY-jk}), (\ref{Gamma-LOY-jk-F}),
(\ref{Gs(l)}). All this has an effect on the real, measurable
values of parameters describing the considered system. The BS
relation in its original form (\ref{BS-1}) and also  the LOY
treatment of this problem are unable to correctly reflect all
complicated processes of this kind.


\begin{thebibliography}{10}
\bibitem{Bell} J. S. Bell and J. Steinberger, in:{\em  "Oxford Int. Conf.
on Elementary Particles 19/25 September 1965:  Proceedings"},
Eds. T. R. Walsh, A. E. Taylor, R.  G.  Moorhouse  and  B.
Southworth, (Rutheford High Energy Lab., Chilton, Didicot 1966),
pp.  195  --- 222.
\bibitem{Horwitz} L. P. Horwitz, J. P. Marchand, {\em Helv. Phys. Acta},
{\bf 42}, (1969), 801.
\bibitem{Barmin} V. V. Barmin, et al.,  {\em
Nucl. Phys.}, {\bf B247}, (1984) 293.
\bibitem{Comins}
E. D. Comins and P. H. Bucksbaum, {\em Weak interactions of
Leptons and Quarks}, (Cambridge University Press, 1983).
\bibitem{dafne1}
L. Maiani, in {\em The Da$\Phi$ne Physics Handbook}, vol. 1, Eds.
L. Maiani, G. Pancheri and N. Paver, SIS
--- Pubblicazioni, INFN  --- LNF, Frascati, 1992; pp. 21 --- 44.
V. S. Demidov, E. Shabalin, ibidem, pp. 45 --- 86.
\bibitem{dafne2}
L. Maiani,  in {\em The Second Da$\Phi$ne Physics Handbook}, vol.
1, Eds. L. Maiani, G. Pancheri and N. Paver, SIS
--- Pubblicazioni, INFN  --- LNF, Frascati, 1995; pp. 3 --- 26.
\bibitem{Bigi}
I. I. Bigi and A. I. Sanda, {\em CP Violation}, Cambridge
University Press, Cambridge, 2001.
\bibitem{leonid1}
L. A. Khalfin, Preprint of the University of Texas at Austin: New
Results on the CP--violation problem,  (Report DOE--ER40200-211,
Feb. 1990).
\bibitem{leonid2}
L. A. Khalfin,   {\em Foundations of Physics}, {\bf 27}, (1997),
1549 and references one can find therein.
\bibitem{Apostolakis}
A. Apostolakis et al., CPLEAR Collaboration, Phys. Lett., {\bf B
456}, (1999), 297.
\bibitem{Rouge}
A. Rouge, hep--ph/9909205.
\bibitem{Tsai1}
K. Kojima, et al, {\em Progr. Theor. Phys.}, {\bf 95}, (1996),
913.
\bibitem{Tsai2}
K. Kojima, et al, {\em Progr. Theor. Phys.}, {\bf 97}, (1997),
103.
\bibitem{Tsai3}
Y. Takeuchi, S. Y. Tsai, hep--ph/0208148.
\bibitem{Hayakawa}
M. Hayakwa, hep-ph/9704418.
\bibitem{Lee1}
T. D. Lee, R. Oehme  and  C.  N.  Yang,  {\em Phys. Rev.}, {\bf
106}, (1957) 340.
\bibitem{Lee2}
T. D. Lee and C. S.  Wu,  {\em Annual Review of Nuclear Science,}
{\bf 16}, (1966) 511.
\bibitem{LOY1}
J. W. Cronin, {\em Rev. Mod. Phys.}, {\bf 53}, (1981) 373.
\bibitem{ww}
V. F. Weisskopf and E. T. Wigner, {\em Z. Phys.}, {\bf 63} (1930)
54.
\bibitem{plb-2002}
K. Urbanowski,  {\em Phys. Lett.}, {\bf B 540}, (2002), 89 and
hep--ph/0201272;
\bibitem{Khalfin}
L. A. Khalfin, Zh. Eks. Teor. Fiz., {\bf 33}, (1957), 1371, [in
Russian]; Sov. Phys.--JETP, {\bf 6}, (19588, 1053.
\bibitem{ur98}K. Urbanowski,
{\em Int. J. Mod. Phys.} {\bf A 13}, (1998), 965.
\bibitem{ijmpa93}
K. Urbanowski, Int. J. Mod.  Phys.  {\bf  A 8},  (1993) 3721.
\bibitem{improved}
K. Urbanowski and J. Piskorski, {\em Found. Phys.}, {\bf 30},
(2000), 839.
\bibitem{Pi00} J. Piskorski, {\em Acta Phys. Polon.}, {\bf B 31}, (2000),
773.
\bibitem{Pi03} J. Piskorski, {\em Acta Phys. Polon.}, {\bf B 37}, (2003),
31.
\bibitem{bull} K. Urbanowski, Bull. de L'Acad. Polon. Sci.: Ser. sci.
phys. astron., {\bf 27}, (1979), 155.
\bibitem{acta}
K. Urbanowski, Acta Phys. Polon. {\bf B 14} (1983) 485.
\bibitem{plb}
K. Urbanowski, Phys. Lett. {\bf B 313}, (1993) 374.
\bibitem{pra}
K. Urbanowski, Phys. Rev. {\bf A 50}, (1994) 2847.
\bibitem{ijmpa95}
K. Urbanowski, Int. J. Mod. Phys., {\bf A10}, (1995), 1151.
\bibitem{KR}
W. Kr\'{o}likowski and J. Rzewuski, Bull. Acad. Polon. Sci. {\bf
4} (1956) 19. W. Kr\'{o}likowski and J. Rzewuski, Nuovo. Cim. {\bf
B 25} (1975) 739 and references therein.
\bibitem{chiu}
C. B. Chiu and E. C. G. Sudarshan, Phys. Rev. {\bf D 42} (1990)
3712; E. C. G. Sudarshan, C. B. Chiu and G. Bhamathi, Unstable
Systems in Generalized Quantum Theory, Preprint DOE-40757-023 and
CPP-93-23, University of Texas, October 1993.


\end{thebibliography}
\end{document}